\documentclass[a4paper,12pt]{article}
\usepackage{amssymb}

\usepackage{epsfig}
\usepackage{latexsym}

\begin{document}


\begin{center}
{\huge \textit{k}-Inflation\\}
\bigskip
{\large C. Armend\'ariz-Pic\'on\raisebox{1ex}{\footnotesize\,a)},
T. Damour\raisebox{1ex}{\footnotesize\,b), c)},
V. Mukhanov\raisebox{1ex}{\footnotesize\,a)}\\}
\medskip
\raisebox{1ex}{\scriptsize a)} Ludwig-Maximilians-Universit\"at,
Sektion Physik \\ M\"unchen, Germany \\ 
{\tt \small http://www.muk.physik.uni-muenchen.de\\}
\medskip
\raisebox{1ex}{\scriptsize b)}
Institut des Hautes \'Etudes Scientifiques
\\ Bures sur Yvette, France \\
{\tt \small http://www.ihes.fr\\}
\medskip
\raisebox{1ex}{\scriptsize c)}
Gravity Probe B, Hansen Experimental Physics Laboratory\\
Stanford University, Stanford, U.S.A. \\
{\tt \small http://einstein.stanford.edu}
\end{center}

\begin{abstract}
It is shown that a large class of higher-order (i.e. non-quadratic) scalar 
kinetic terms can,
without the help of potential terms, drive an inflationary evolution
starting from rather generic initial conditions. In many models, this
kinetically driven inflation (or ``$k$-inflation'' for short) rolls slowly from 
a high-curvature initial
phase, down to a low-curvature phase and can exit inflation to end up being
radiation-dominated, in a naturally graceful manner. We hope that this novel
inflation mechanism might be useful in suggesting new ways of reconciling
the string dilaton with inflation.
\end{abstract}


\section{Introduction}
\setcounter{equation}{0}

The inflationary paradigm \cite{L90} offers the attractive possibility of
resolving many of the puzzles of standard hot big bang cosmology. The
crucial ingredient of most successful inflationary scenarios is a period of
``slow-roll'' evolution of a scalar field $\varphi $ (the ``inflaton''),
during which the potential energy $V(\varphi )$ stored in $\varphi $
dominates its kinetic energy $\dot{\varphi}^{2}/2$ and drives a
quasi-exponential expansion of the Universe. At present there exists no
preferred concrete inflationary scenario based on a convincing realistic
particle physics model. In particular, though string theory provides one
with several very weakly coupled scalar fields (the moduli), which could be
natural inflaton candidates, their non-perturbative potentials $V(\varphi )$ do 
not seem fit to sustain slow-roll inflation because, for large
values of $\varphi $, they tend either to grow, or to tend to zero, too fast. It
is therefore important to explore novel possibilities for implementing an
inflationary evolution of the early Universe.

The aim of this work is to point out that, even in absence of any potential
energy term, a general class of non-standard (i.e. non-quadratic) 
\textit{kinetic-energy} terms 
$\mathcal{L}(\dot{\varphi})$, for a scalar field $\varphi $, can drive an
inflationary evolution of the same type as the usually considered
potential driven inflation. By ``usual type of inflation'' we mean here an
accelerated \textit{expansion} (in the Einstein conformal frame) during
which the curvature scale starts around a Planckian value and then \textit{%
decreases} monotonously. By contrast, the pre-big bang scenario \cite{PBB}
uses a \textit{standard} (quadratic) kinetic-energy term $\dot{\varphi}^{2}/2$ 
to drive
an accelerated \textit{contraction} (in the Einstein frame) during which the
curvature scale \textit{increases}. Though we shall motivate below the
consideration of non-standard kinetic terms by appealing to the existence,
in string theory, of higher-order corrections to the effective action for
$\varphi $, we do not claim that the structure needed for implementing our
``kinetically driven'' inflation arises inevitably in string theory. The aim
of this work is more modest. We draw attention to a new mechanism for
implementing inflation. A large class of the (toy) models we shall consider
satisfy two of the most crucial requirements of inflationary scenarios: (i)
the scalar perturbations are ``well-behaved'' during the  inflationary stage, 
and (ii) there exist natural mechanisms for exiting inflation in a ``graceful''
manner. We leave to future work a more detailed analysis of the
observational consequences of our kinetically driven inflation (spectrum of
scalar and tensor perturbations, reheating, \ldots). We hope that,
by enlarging the basic ``tool-kit'' of inflationary cosmology, our mechanism
could help to locate which sector of string-theory (if any) has inflated a
strongly curved initial state into our presently observed large and weakly
curved Universe.


\section{General Model}
\label{sec:general-model}
\setcounter{equation}{0}

We consider a single scalar field $\varphi $ interacting with gravity
through non-standard kinetic terms, 
\begin{equation}
  \label{eq:einstein-action}
  S=\int d^{4}x\,\sqrt{g}\left[ -\frac{R}{6\kappa ^{2}}+p(\varphi ,\nabla
    \varphi )\right] .  
\end{equation}
Here, $\kappa ^{2}\equiv 8\pi G/3$ and we use the signature $(+---)$. For
simplicity, we only consider Lagrangians $p$ which are local functions of $%
\nabla _{\mu }\varphi $ and therefore, depend only on the scalar 
\begin{equation}
  X\equiv \frac{1}{2}(\nabla \varphi )^{2}.
\end{equation}
As we bar here the consideration of potential terms, we must impose that the
function $p(\varphi,X)$ vanish when $X\rightarrow 0$. Near $X=0$, a
generic kinetic Lagrangian $p(\varphi,X)$ is expected to admit an expansion
of the form 
\begin{equation}
  \label{eq:p-expansion}
  p(\varphi ,X)=K(\varphi )X+L(\varphi )X^{2}+\cdots.  
\end{equation}
One of the possible particle theory motivations for taking this kind of
Lagrangian seriously could be the following.  Let us consider only gravity 
and some moduli field $\varphi $ (which could be the dilaton, or some other
moduli) in string theory. It is well-known that $\alpha^{\prime}$ corrections
(due to the massive modes of the string) generate a series of higher-derivative 
terms in the low-energy effective action $S_{\rm eff}$, while string-loop 
corrections generate a non-trivial 
moduli dependence of the coefficients of the various kinetic terms. This leads
to a structure of the type (in the string frame $\hat{g}_{\mu\nu}$) 
\begin{eqnarray}
  \label{eq:string-effective-action}
  S_{\rm eff} &=&\frac{1}{6\kappa^{2}}\int d^{4}x\sqrt{\hat{g}}\Big\{
  -B_{g}(\varphi)\hat{R}-B_{\varphi}^{(0)}(\hat{\nabla}\varphi)^{2} \\
  &&+\alpha^{\prime}\left[c_{1}^{(1)}B_{\varphi}^{(1)}(\varphi)(\hat{
      \nabla}\varphi)^{4}+\cdots\right] +O(\alpha'\,^{2})\Big\}, 
  \nonumber
\end{eqnarray}
where the ellipsis stands for other four-derivative terms (like $(\Box
\varphi)^{2},\hat{R}_{\mu\nu\rho\sigma}^{2}, \ldots )$. In the case
where $\varphi $ is the dilaton the coupling functions are of the form 
\begin{eqnarray*}
  B_{g}(\varphi ) &=&e^{-\varphi }+c_{g,0}+c_{g,1}e^{\varphi }+\cdots , \\
  B_{\varphi }^{(0)}(\varphi ) &=&e^{-\varphi }+c_{\varphi ,0}+c_{\varphi
    ,1}e^{\varphi }+\cdots , \\
  B_{\varphi }^{(1)}(\varphi ) &=&e^{-\varphi }+\cdots ,
\end{eqnarray*}
where the ellipsis contain higher contributions in $g_{\rm string}^{2}=
e^\varphi$, including non-perturbative ones. Transforming Eq.
(\ref{eq:string-effective-action}) to the Einstein frame $g_{\mu\nu
}=B_{g}(\varphi)\hat{g}_{\mu\nu}$, and neglecting the other possible four
derivative terms in Eq. (\ref{eq:string-effective-action}), leads to the
effective action (\ref{eq:einstein-action}) with $\varphi$-kinetic terms of 
the form (\ref{eq:p-expansion}) where 
\begin{eqnarray}
  K(\varphi ) &=&3\frac{B_{g}^{\prime }{}^{2}(\varphi )}{B_{g}^{2}(\varphi )}-2
  \frac{B_{\varphi }^{(0)}(\varphi )}{B_{g}(\varphi )}, \\
  L(\varphi ) &=&c_{1}^{(1)}\frac{2\alpha ^{\prime }}{3\kappa ^{2}}B_{\varphi
    }^{(1)}(\varphi ).
\end{eqnarray}
In the case where $\varphi $ is the dilaton (so that $g_s=e^{\varphi /2}$
is the string coupling), we have, in the weak-coupling limit
$g_s^{2}=e^{\varphi}\ll 1$, $B_{g}(\varphi )\simeq B_{\varphi}^{(0)}
(\varphi)\simeq B_{\varphi}^{(1)}(\varphi)\simeq e^{-\varphi}$ so
that $K(\varphi)\simeq 1$ and $L(\varphi)\propto e^{-\varphi}$.
However, when $g_s$ becomes of the order of unity it is not a priori
excluded that $\ K(\varphi)$ and $L(\varphi)$ could become more
complicated functions of $\varphi$. We shall give below some examples of such 
possible complicated behaviours.

Returning now to a general Lagrangian $p(\varphi ,X)$ the ``matter''
energy-momentum tensor reads 
\begin{equation}
  \label{eq:energy-momentum-tensor}
  T_{\mu \nu }\equiv \frac{2}{\sqrt{g}}\frac{\delta S_{\varphi }}{\delta
    g^{\mu \nu }}=\frac{\partial p(\varphi ,X)}{\partial X}\nabla _{\mu }\varphi
  \nabla _{\nu }\varphi -p(\varphi ,X)g_{\mu \nu }.
\end{equation}
Equation (\ref{eq:energy-momentum-tensor}) shows that, if $\nabla _{\mu
}\varphi $ is time-like (i.e. $X>0$), our scalar field action is
``equivalent'' to a perfect fluid ($T_{\mu \nu }=(\varepsilon +p)u_{\mu
}u_{\nu }-p\,g_{\mu \nu }$) with pressure 
\begin{equation}
  \label{eq:pressure}
  p=p(\varphi ,X),  
\end{equation}
energy density 
\begin{equation}
  \label{eq:energy-density}
  \varepsilon =\varepsilon (\varphi ,X)\equiv 2X\frac{\partial p(\varphi ,X)}
  {\partial X}-p(\varphi ,X),  
\end{equation}
and four-velocity 
\begin{equation}
  \label{eq:four-velocity}
  u_{\mu }=\sigma \frac{\nabla _{\mu }\varphi }{\sqrt{2X}},
\end{equation}
where $\sigma $ denotes the sign of $\dot{\varphi}=\nabla _{0}\varphi $.

As usual, in inflationary cosmology, we consider a flat background Friedmann
model $ds^2=dt^2-a^2(t)d\mathbf{x}^2$ and an homogeneous background scalar
field $X=\frac{1}{2}\dot{\varphi}^2$. A convenient minimal set of
independent evolution equations for $a(t)$ and $\varphi(t)$ is (with $H\equiv%
\dot{a}/a$) 
\begin{eqnarray}
\label{eq:Friedmann}
H^2 &=&\kappa^2 \varepsilon \, , \\
\label{eq:energy-conservation}
\dot{\varepsilon}&=&-3H(\varepsilon+p).
\end{eqnarray}
It will be also useful to refer to other (redundant) forms of the evolution
equations: 
\begin{eqnarray}  
\frac{\ddot{a}}{a}&=&-\frac{1}{2}\kappa^2(\varepsilon+3p) \, , \\
\label{eq:canonical}
\frac{1}{a^3}\frac{d}{dt}(a^3\pi)&=&\dot{\pi}+3H\pi=\frac{\partial p}
{\partial \varphi},
\end{eqnarray}
where $\pi\equiv\partial p/\partial\dot{\varphi}=\dot{\varphi}\,\partial
p/\partial X$ denotes the momentum conjugate to $\varphi$. [Note that
equation (\ref{eq:energy-density}) reads $\varepsilon=\dot{\varphi}\partial
p/\partial \dot{\varphi}-p$, which is the usual energy associated to the
Lagrangian $p(\varphi,\dot{\varphi})$.]

In this work we shall only consider solutions of Eqs. (\ref{eq:Friedmann})-(%
\ref{eq:canonical}) which describe an expanding Universe (in Einstein
frame), that is $H>0.$ This reduces the evolution equations (\ref
{eq:Friedmann}), (\ref{eq:energy-conservation}) to the master equation 
\begin{equation}
\label{eq:master}
\dot{\varepsilon}=-3\sqrt{\varepsilon }(\varepsilon +p).  \label{master}
\end{equation}
Here, and in the following, we use units such that $\kappa ^{2}\equiv 8\pi
G/3=1$. Note that, from Eq. (\ref{eq:Friedmann}), $\varepsilon $ was
constrained to be positive.


\section{Kinetically Driven Inflation - Basic Idea}
\label{sec:k-inflation}
\setcounter{equation}{0}

\begin{figure}[!t]
  \begin{center}
    \epsfig{file=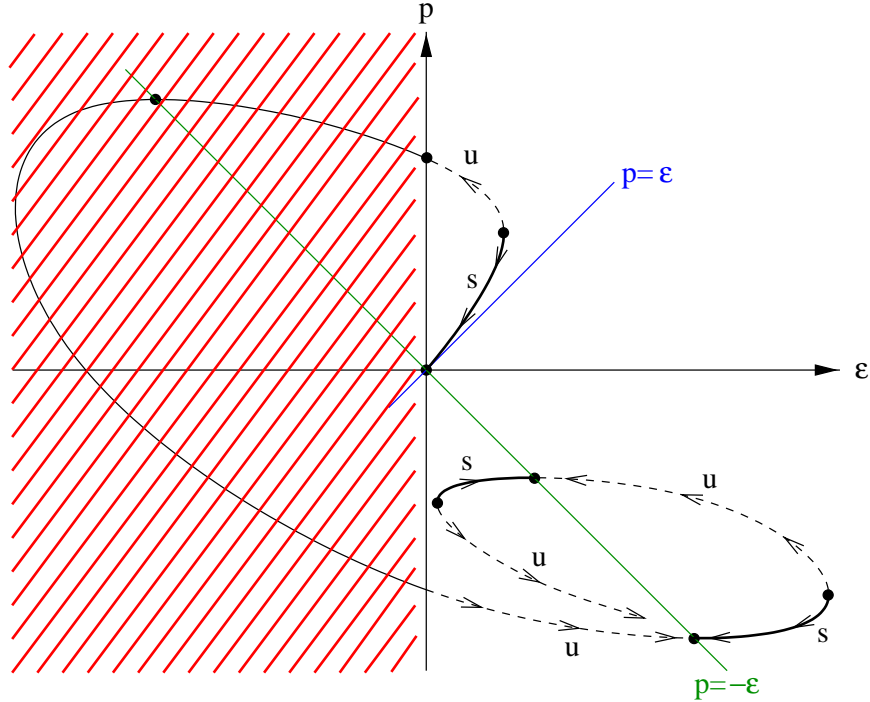}
  \end{center}
  \caption{Graph of the equation of state linking $p$ to $\protect\varepsilon$
for an hypothetical general kinetic Lagrangian $p(\dot{\protect\varphi})$.
The evolution for expanding, flat cosmologies proceeds along the indicated
arrows. The shaded region ($\protect\varepsilon<0$) is excluded. Except for
the origin and the point above it on the vertical axis, the attractors of
the evolution are inflationary fixed points with $p=-\protect\varepsilon$.}
\label{fig:state}
\end{figure}

As a warm up, let us first consider the case where the Lagrangian $p$
depends only on $X=\frac{1}{2}(\nabla \varphi )^{2}$, and not on $\varphi $: 
$p=p(X)$. From Eq. (\ref{eq:energy-density}), the energy-density depends
also only on $X$: $\varepsilon (X)=2X\partial p/\partial X-p(X)$. In
hydrodynamical language it means that we have here an ``isentropic'' fluid,
with a general equation of state relating $p$ to $\varepsilon $: $%
p=f(\varepsilon )$. The master evolution equation can then be qualitatively
solved by looking at the graph of the equation of state $p=f(\varepsilon )$.
The shape of this graph depends very much on the shape of the function
$p=p(X)$, or better, on the shape of the function $p=p(\psi)$ where $\psi
\equiv \dot{\varphi}$, so that $X=\frac{1}{2}\psi^{2}$. Indeed, the
relation 
\begin{equation}
  \varepsilon =\psi \partial p/\partial \psi -p  \label{eq:varepsilon-psi}
\end{equation}
shows that $\varepsilon $ can be read geometrically off the graph
$p=p(\psi)$. More precisely $-\varepsilon$ is the ``intercept''of the
tangent to the curve $p=p(\psi)$ at the point $\psi$, i.e. its
intersection with the vertical axis $\psi=0$. The minus sign means that
$\varepsilon $ is positive (negative) if the tangent intersects the vertical
axis below (above) the $p=0$ horizontal axis. Therefore, if the function 
$p(\psi )=\frac{1}{2}K\psi ^{2}+\frac{1}{4}L\psi ^{4}+\cdots $ is, for
instance, always \textit{convex}, $\partial ^{2}p/\partial \psi ^{2}>0$, as
will be the case if all the coefficients $K,L,\ldots $ appearing in the
expansion (\ref{eq:p-expansion}) are positive, $p$ and $\varepsilon $ will
be always positive. In such a case, Eq. (\ref{eq:master}) shows that
$\varepsilon $ will monotonically decrease towards zero, and the evolution
will be driven to the ``attracting solution'' 
\begin{equation}
  \label{eq:vacuum-attractor}
  \varepsilon =H^{2}\approx \frac{1}{9t^{2}}\quad ,\quad a\approx 
a_{0}\,t^{1/3}.
\end{equation}
This attractor corresponds to the asymptotic equation of state $p\approx
\varepsilon $ valid near $\varepsilon =0$ where the usual kinetic term $%
p\simeq \frac{1}{2}K\psi ^{2}$ dominates. On the other hand, if the function 
$p(\psi )$ is non-convex and has some oscillatory behaviour as $\psi $
increases (i.e., if we consider the general case where the expansion
coefficients $K,L,\ldots $ in Eq. (\ref{eq:p-expansion}) may take negative
values) the graph $p=f(\varepsilon )$ can be more complicated and can allow
for exponential-type inflationary behaviour. Let us first note that, because
of Eq. (\ref{eq:varepsilon-psi}), the extrema of the function $p=p(\psi )$
(or $p=p(X)$) correspond to values where $p=-\varepsilon $, i.e. to \textit{%
fixed points} of the master evolution equation (\ref{eq:master}). For a
general function $p=p(X)$ the graph of the (multiform) equation of state
might resemble Fig.\ref{fig:state}.

From the master Eq. (\ref{eq:master}) \ it follows that $\varepsilon $ will
decrease above the line $p=-\varepsilon $, and increase below it. Fig. \ref
{fig:state} then shows that all the intersection points with the $%
p=-\varepsilon $ line are \textit{attractors} of the (future) evolution. The
arrows in Fig. \ref{fig:state} indicate the evolutive flow, which reverses
(along the graph) at the extrema of $\varepsilon $. The region where $%
\varepsilon <0$ is excluded because it cannot be reached by flat cosmologies
(see Eq. (\ref{eq:Friedmann})). The fixed points lying at the $%
p=-\varepsilon $ line correspond to an exponential inflation 
\begin{equation}
  \label{eq:inflationary-attractor}
  H_{\rm att}^2=\varepsilon_{\rm fixed}\quad ,\quad
  a_{\rm att}(t)=a_{0}\exp\left(\sqrt{\varepsilon_{\rm fixed}}t\right)
\end{equation}
Apart from these inflationary attractors, there are two other attractors (in
the case depicted in Fig. \ref{fig:state}): (i) the origin, where the
evolution is driven toward the solution (\ref{eq:vacuum-attractor})
(corresponding to the ``hard'' equation of state $p=\varepsilon $), and (ii)
the point above the origin on the vertical axis. As we will see later the latter
point lies in the region of  absolute instability and
has therefore no physical significance.

In this work we shall focus on the inflationary attractors
(\ref{eq:inflationary-attractor}) because they exhibit the novel possibility of
getting, for a large set of initial conditions, quasi-exponential (or
power law) inflation out of a purely kinetic Lagrangian. Note that the
condition for the existence of these inflationary attractors can also be
seen in Eq. (\ref{eq:canonical}). In absence of a $\varphi $ dependence, Eq.
(\ref{eq:canonical}) says that $a^{3}\pi $ is constant so that $\pi $ is
attracted toward zero. As the momentum $\pi=\partial p/\partial \dot{\varphi}=%
\dot{\varphi}\,\partial p/\partial X$, we see that non trivial ($\dot{\varphi%
}\neq 0$) attractors can exist if the kinetic terms are non-standard so that 
$\partial p/\partial X$ can vanish for non zero values of $\dot{\varphi}$ .
The extremal values of $p$ correspond to the inflationary attractors
discussed above.

The labels ``s'' and ``u'' in Fig. \ref{fig:state} (which stand for
``stable'' and ``unstable'') indicate whether, for our present isentropic
equation of state $p=f(\varepsilon)$, the squared speed of sound
$c_s^2=dp/d\varepsilon$ is positive or negative, respectively. This issue
will be further discussed below.

Note that the ``price'' to pay for having inflationary attractors as in Fig.
\ref{fig:state} is the existence of regions, in phase space $(\varphi ,%
\dot{\varphi})$, with negative energy density. We shall assume in this work
that this is not physically forbidden. All the cosmological evolutions that
we shall consider below stay always in the positive energy regions, and the
existence, elsewhere in phase space, of negative $\varepsilon $ regions does
not necessarily cause some instabilities along our evolutionary tracks.

The simple case of a kinetic Lagrangian depending only on 
$\dot{\varphi}$ considered above is the analog, for kinetically driven inflation
(or ``\textit{k}-inflation'' for short), of a de Sitter model 
with constant energy density. It is clear that both models should have similar 
problems. Namely, there is no natural graceful exit, no smooth transition to a 
Friedmann Universe and the cosmological perturbations are ``ill-defined''.
To avoid these problems we should allow the coefficients in the expansion 
(\ref{eq:p-expansion}) of the Lagrangian $p$ to depend on the scalar field
$\varphi$.


\section{``Slow-Roll'' \textit{k}-Inflation}
\label{sec:slow-roll-inflation}
\setcounter{equation}{0}

The simplest way to realize successfully the idea of \textit{k}-inflation is to
consider the analog of ``slow-roll'' potential driven inflation, in which the
potential $V(\varphi )$ in the Lagrangian $\mathcal{L}(\varphi ,\dot{\varphi}%
)=\frac{1}{2}\dot{\varphi}^{2}-V(\varphi )$ dominates the kinetic term $\dot{%
\varphi}^{2}/2$ and evolves slowly. For the concept of \textit{k}-inflation
to have a relevance to a large class of models, we need to consider a
general kinetic Lagrangian $p(\varphi ,X)$. The idea is therefore to find
the conditions under which the influence of the non-trivial $\varphi $
dependence of $p(\varphi ,X)$ will represent only a relatively small
perturbation of the attraction toward exponential inflation discussed in
Section \ref{sec:k-inflation}. To do that in a concrete manner it is
convenient to focus henceforth on the simplest kinetic Lagrangian,
containing only $\dot{\varphi}^{2}$ and $\dot{\varphi}^{4}$ terms, namely 
\begin{equation}
\label{eq:simplest-p-expansion}
  p(\varphi ,X)=K(\varphi )X+L(\varphi )X^{2}=\frac{1}{2}K(\varphi )(\nabla
  \varphi )^{2}+\frac{1}{4}L(\varphi )(\nabla \varphi )^{4}.
\end{equation}
Let us first motivate the possibility of rather arbitrary functions $%
K(\varphi ),L(\varphi )$ by considering again the low-energy effective action of
string theory (\ref{eq:string-effective-action}). As we mentioned above, in the 
weak coupling
limit $K(\varphi )\simeq 1$ and $L(\varphi )\propto e^{-\varphi }.$
However, when $g_{s}$ becomes of order unity it is not a priori excluded
that $K(\varphi )$ could change sign. For instance we could consider a
simple model (of the type considered in Refs. \cite{DP94}, \cite{DV96}),
where the coupling functions $B(\varphi)$ in the action 
(\ref{eq:string-effective-action}) are the 
same, that is
$B_{g}(\varphi)=B_{\varphi}^{(0)}(\varphi)=B_{\varphi}^{(1)}(\varphi)=B(\varphi)
$. In this model 
\begin{equation}
  K(\varphi )=3\left(\frac{B'(\varphi)}{B(\varphi)}\right)^2-2
\end{equation}
might become negative when $g_{s}$ reaches values of order unity. In fact a
model, incorporating string loop corrections, considered in Ref. \cite{FMS99}
has 
\begin{equation}
  \label{eq:FMS-K}
  K(\varphi)=1-3k\,e^{\varphi}\frac{6+k\,e^{\varphi}}{(3+k\,e^{\varphi})^2},
\end{equation}
where $k=3\delta^{\rm GS}/8\pi ^{2}$ is a positive parameter. The R.H.S. of Eq.
(\ref{eq:FMS-K}) becomes negative when $e^{\varphi}>3(\sqrt{6}-2)/2k$.
Motivated by these examples, we shall assume, as simplest toy model exhibiting
interesting dynamics, a Lagrangian of the form (\ref{eq:simplest-p-expansion}%
) with a function $K(\varphi)$ which is positive in some range of values of 
$\varphi $ (``weak-coupling-domain'') and becomes negative in some other
range (``strong-coupling-domain''). On the other hand, to ensure the
positivity of $\varepsilon $ for large field gradients $X$, we shall assume
that the function $L(\varphi )$ remains always positive.

The equation of state in the model (\ref{eq:simplest-p-expansion}) is
parametrically given by 
\begin{eqnarray}
  p&=&K(\varphi)X+L(\varphi)X^2, \\
  \varepsilon&=&K(\varphi)X+3L(\varphi)X^2.
\end{eqnarray}

\begin{figure}[!b]
  \begin{center}
    \epsfig{file=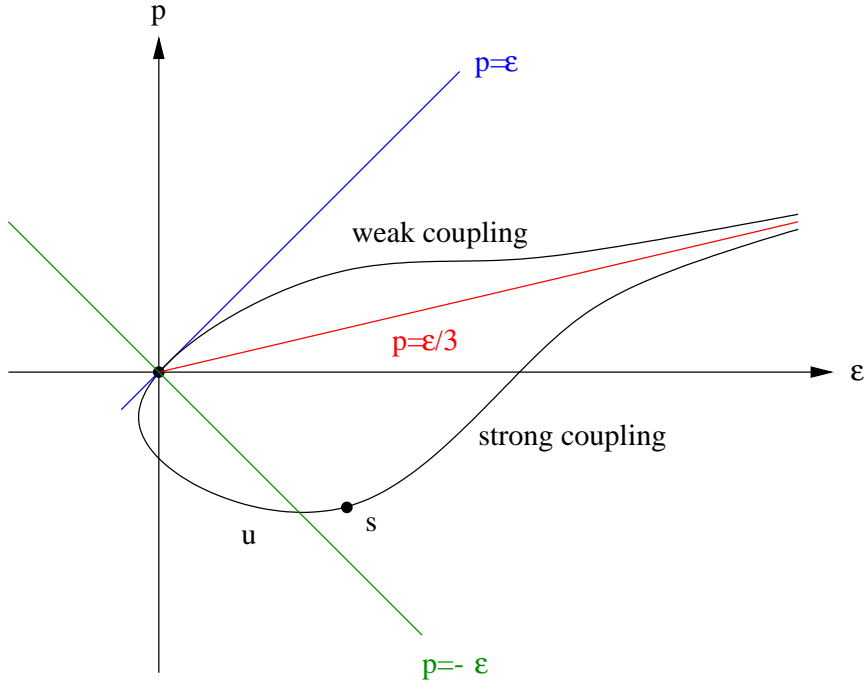}
  \end{center}
  \caption{Change of form of the equation of state $p=f(\protect\varepsilon,
    \protect\varphi)$ in the model (\ref{eq:simplest-p-expansion}) as
    $\protect\varphi$ varies from the weak-coupling region
    ($K(\protect\varphi)>0$) to
    the strong-coupling one ($K(\protect\varphi)<0$).}
  \label{fig:model-state}
\end{figure}

We represent in Fig. \ref{fig:model-state} the change in the form of the
equation of state $p=f(\varepsilon,\varphi)$ as $\varphi$ varies from the
weak-coupling region ($K>0$) to the strong-coupling one ($K<0$). Note that for 
large
values of $X$, the equation of state asymptotes the one of radiation
($p=\varepsilon/3$), while it is tangent to the hard equation of state
($p=\varepsilon$; with $\mathrm{sgn}(\varepsilon)=\mathrm{sgn}(K)$) when
$X\to0 $. In the strong coupling domain there appears (in the adiabatic
approximation where $\varphi$ is treated as constant) an inflationary fixed
point where $p_{\rm fixed}=-\varepsilon_{\rm fixed}$.

Let us investigate under what conditions on the functions $K(\varphi )$ and
$L(\varphi)$ one can indeed approximately solve the master evolution
equation (\ref{eq:master}) by considering that the variation of $\varphi$
brings only a small perturbation onto the simple $\varphi $-independent
evolution studied above. We shall simplify this study by redefining the
scalar field and working with the new field variable $\varphi^{\rm new}=\int
d\varphi^{\rm old}L^{1/4}(\varphi^{\rm old})$ (which is well defined because we
assume $L(\varphi)>0$). With this definition $L_{\rm new}(\varphi^{\rm new})=1$
and $K_{\rm new}(\varphi^{\rm new})=K_{\rm old}(\varphi^{\rm old})/L_{\rm old}^
{1/2}(\varphi^{\rm old})$. In other words, we can assume, without loss of 
generality, that $L(\varphi)=1$. With this simplification the zeroth-order 
``slow-roll'', or ``adiabatic'', solution to Eq. (\ref{eq:master}), i.e. the
instantaneous attractive fixed point of Eq. (\ref{eq:master}) (solution of
$0=\varepsilon+p=2X\partial p/\partial X)$ corresponds to 
\begin{eqnarray}
  \label{eq:X-slow-roll}
  X_{\rm sr} &=&\frac{1}{2}\overline{K}(\varphi_{\rm sr}), \\
  \dot{\varphi}_{\rm sr} &=&\sigma\sqrt{\overline{K}(\varphi_{\rm sr})}, \\
  \varepsilon_{\rm sr} &=&\frac{1}{4}\overline{K}^{2}(\varphi_{\rm sr}), \\
  \label{eq:H-slow-roll}
  H_{\rm sr} &=&\frac{1}{2}\overline{K}(\varphi_{\rm sr}).
\end{eqnarray}
Here $\overline{K}(\varphi)\equiv -K(\varphi)$ is positive (in the slow-roll
domain) and $\sigma $ denotes the sign of $\dot{\varphi}$. The time
evolution of the slow-roll \textit{k}-inflation is given, from Eqs. (\ref
{eq:X-slow-roll})-(\ref{eq:H-slow-roll}), by simple quadratures (the
subscript ``in'' denotes initial values) 
\begin{eqnarray}
  t-t_{\rm in} &=&\sigma \int_{\varphi_{\rm in}}^{\varphi}\frac{d\varphi}
  {\sqrt{\overline{K}(\varphi)}}, \\
  N &\equiv &\ln \left( \frac{a(t)}{a_{\rm in}}\right)=\frac{\sigma}{2}
  \int_{\varphi_{\rm in}}^{\varphi}\sqrt{\overline{K}(\varphi)}d\varphi .
\end{eqnarray}
[The notation $N$ is introduced to denote the number of $e$-folds of
inflation.]

The post-slow-roll approximation, $X=X_{\rm sr}+\delta {X}$, is then obtained
by rewriting the master equation (\ref{eq:master}) as 
\begin{equation}
  \frac{\partial p}{\partial X}=-\overline{K}+2X=2\delta 
X=-\frac{\dot{\varepsilon}}
  {6X\sqrt{\varepsilon}},
\end{equation}
and replacing the slow-roll approximations (\ref{eq:X-slow-roll})-(\ref
{eq:H-slow-roll}) in the R.H.S. This yields 
\begin{equation}
  \label{eq:slow-roll-parameter}
  \frac{\delta X}{X_{\rm sr}}\simeq -\frac{1}{12}\frac{\dot{\varepsilon}}
  {\varepsilon ^{3/2}}\simeq -\frac{\sigma}{3}\frac{\overline{K}^{\prime}}
  {\overline{K}^{3/2}}\simeq +\frac{2\sigma 
}{3}\left(\frac{1}{\sqrt{\overline{K}}}
  \right)',  
\end{equation}
where the prime denotes a derivative with respect to $\varphi $. The criterion
for the validity of our previous slow-roll solution
(\ref{eq:X-slow-roll})-(\ref{eq:H-slow-roll}) is 
\begin{equation}
  \label{eq:slow-roll-condition}
  \frac{\delta X}{X}\ll 1,  
\end{equation}
i.e. $(\overline{K}^{-1/2})^{\prime}\ll 3/2$ (when keeping $L(\varphi)\neq 1$ it 
would read $L^{-1/4}\partial (L^{1/4}\overline{K}^{-1/2})$ $/\partial\varphi \ll 
3/2$).
This condition is as easily satisfied as the usual slow-roll condition
for potential driven inflation. Examples of functions $\overline{K}(\varphi)$ 
that
satisfy this condition are: (i) any power law or exponential (or 
super-exponential) growth as $\varphi\rightarrow\infty$, (ii) any levelling 
off of $\overline{K}(\varphi)$ ($\overline{K}(\varphi)\rightarrow {\rm limit}$, 
with
$\overline{K}^{\prime}(\varphi)\rightarrow 0$) as $\varphi\rightarrow\infty $,
or (iii) a sufficiently fast pole-like growth of $\overline{K}(\varphi)\propto
(\varphi_*-\varphi)^{-\alpha }$, with $\alpha>2$, as
$\varphi\rightarrow\varphi_*$.

Note that during slow-roll \textit{k}-inflation the following useful relation
\begin{equation}
  \label{eq:e-plus-p}
  \frac{\varepsilon +p}{\varepsilon }\simeq 4\frac{\delta X}{X_{\rm sr}},
\end{equation}
is satisfied, that is, the fractional compensation of the energy density by
the negative pressure is proportional to the small parameter $\delta
X/X_{sr}.$ Therefore, in those models where $\varepsilon+p> 0,$ or
equivalently, the energy density decreases in the course of expansion, 
$\delta X$ is positive. It is also obvious from (\ref{eq:e-plus-p}), that 
inflation ends when $\delta X/X_{\rm sr}$ becomes of the order of unity, i.e. 
when the slow-roll condition (\ref{eq:slow-roll-condition}) is violated.

For any function $\overline{K}(\varphi)$ (and more generally for any $p(\varphi
,X)$) satisfying the slow-roll criterion we can visualize our
\textit{k}-inflationary behaviour as being one point (i.e. one value of
$X=X_{\rm sr}+\delta X$) on an adiabatically varying equation of state graph
of the type of Figs. {\ref{fig:state}} or \ref{fig:model-state}. [The
adiabatic variation we mention corresponding to the fact that each graph
corresponds to some specific, instantaneous value of $\varphi$ , which is
itself evolving]. Eq. (\ref{eq:slow-roll-parameter}) (or its generalization
to a generic $p(\varphi,X)$) then tells us that the point
$X=X_{\rm sr}+\delta X$ is always \textit{displaced} away from the intersections
of the ($\varphi $-instantaneous) graph with the $p=-\varepsilon$ line.

Overall, the qualitative behaviour of the solutions we are focussing on is
the following: Initially, we start with some representative point in the
$(\varepsilon ,p)$ plane lying on an equation of state graph corresponding to
some initial value of $\varphi $, deep into the strong coupling domain. We
assume that, for strong coupling, the slow-roll criterion is very well
satisfied. In a first evolution stage, we can neglect the $\varphi$-dependence
of the equation of state because there is a fast attraction
taking just a few $e$-folds of the representative point toward the nearest
inflationary attractor (this stage is described by the arrows in Fig. \ref
{fig:state}). After this initial stage, we can consider that our
representative point follows the (post-) slow-roll motion
$X=X_{\rm sr}+\delta X$, corresponding to a representative point near but away
from the $p=-\varepsilon $ line (such a point is indicated in
Fig. \ref{fig:model-state}). As the evolution continues, the slow-roll
condition is less and less well satisfied and the representative point
straggles more and more away from the $p=-\varepsilon $ line. At some point in
the evolution the slow-roll criterion (\ref{eq:slow-roll-condition}) becomes
violated ($\delta X/X\sim 1$) and one naturally exits the inflationary stage.
We shall come back later to this exit mechanism.

\begin{figure}
  \begin{center}
    \epsfig{file=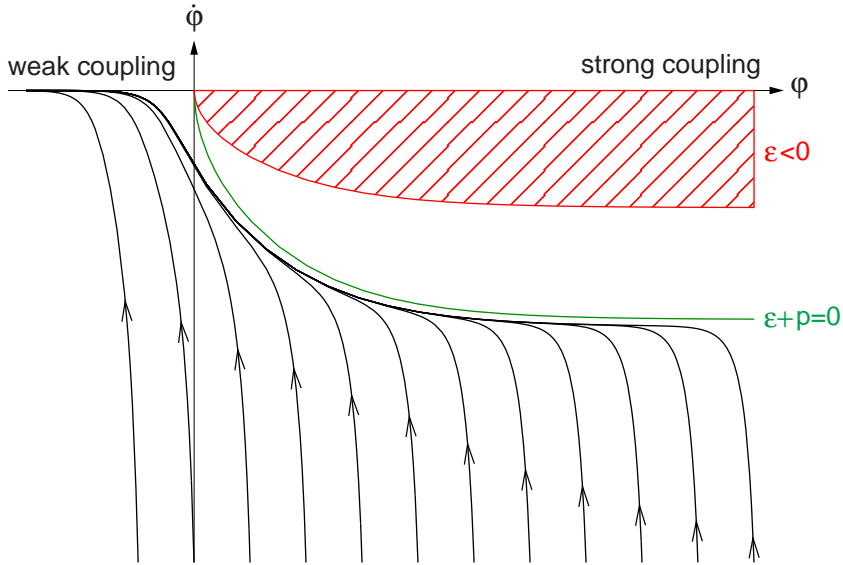}
  \end{center}
  \caption{\label{fig:phase}Schematic phase diagram of ``slow-roll'' 
    $k$-inflation. Trajectories approach the attractor but do not reach the line 
    $\epsilon+p=0$ where the speed of sound vanishes. Around the point  
    where the slow-roll condition is violated, the solutions leave the
    inflationary stage and approach then smoothly the vacuum
    $\dot{\varphi}=0$.}
\end{figure}

The qualitative picture  of the evolution just represented (based on a
succession of graphs in the $(\varepsilon,p)$ plane) can also be globally
visualized by a phase-space picture in the $(\varphi,\dot{\varphi})$ plane
(see Fig. \ref{fig:phase}). ``Slow-roll'' inflation on this graph corresponds
 to the portion of the separatrix (attractor) given by 
\begin{equation}\label{eq:separatrix}
  \dot{\varphi}=\sigma \sqrt{\stackrel{\_}{K}}\left( 1+\frac{\delta X}
    {X_{\rm sr}}+...\right)^{1/2},
\end{equation}
where $\delta X/X\ll 1.$ When $\delta X/X$ becomes of the order of one
a graceful exit from inflation takes place, see Section 
\ref{sec:exit-mechanisms}. The phase diagram Fig. \ref{fig:phase} is very 
similar to 
the one of potential driven slow-roll inflation \cite{BGZK85}.
We see that the set of initial configurations of the scalar field which lead
to inflation has nonzero measure. Therefore the problem of initial conditions
 here is very similar to what one has in the case of chaotic inflation
\cite{L90}. We expect that in analogy to the other models of inflation 
self-reproduction of the Universe \cite{LLM94} can take place in our
\textit{k-}inflationary model. However this question needs a special
investigation and we leave it to future work.

Let us also mention that one can easily build a model in which one starts 
initially in the weak coupling regime, with the field $\varphi$ evolving towards 
the strong coupling regime. The function $K$ can then be arranged in
such a way that the Universe leaves the weak coupling regime to enter an  
inflationary stage and finally exits inflation.


\section{``Power Law'' \textit{k}-Inflation}
\label{sec:power-law-kinflation}
\setcounter{equation}{0}

It is well known that, in the usual potential driven inflationary scenario,
if the potential depends exponentially on the scalar field, there exists an 
attractor solution which describes a power law inflating Universe
(see for instance \cite{LM85}). There is no graceful exit from
inflation if the potential is exponential everywhere. Therefore to solve
the graceful exit problem one should assume that the exponential potential
is a valid approximation of a more realistic and complicated potential
only within some limited range of values of the scalar field $\varphi$. As we 
show
in this section, one can get an analogous power law \textit{k}-inflation within
the class of models which we consider in this paper. 

Let us again consider
the model with the Lagrangian (\ref{eq:simplest-p-expansion}). For the
purposes of this section it is convenient to make a new field redefinition 
(valid only in the region where $K<0$)
and rewrite the Lagrangian (\ref{eq:simplest-p-expansion}) in terms of
the new field variable $\varphi^{\rm new}=\int\sqrt{L(\varphi^{\rm old})/
|K(\varphi^{\rm old})|}\,d\varphi^{\rm old}$. This 
yields 
\begin{equation} 
  \label{eq:new-lagrangian}
  p=f\left(\varphi\right)\left(-X+X^{2}\right),
\end{equation}
where $f\left(\varphi\right)\equiv f(\varphi^{\rm new})=K^{2}(\varphi
^{\rm old})/L(\varphi^{\rm old})$ and $X\equiv X^{\rm new}=
(L/|K|)X^{\rm old}$.

Working with this Lagrangian one can try to find out whether there is a
function $f\left(\varphi\right)$ for which the master equation
(\ref{eq:master}) has an \textit{exact} solution which describes power law
inflation. In the case of power law inflation 
\begin{equation}
  \label{eq:gamma}
  \varepsilon +p=\gamma \varepsilon,
\end{equation}
where $\gamma$ is a constant. Substituting $\varepsilon $ and $p$ into the
last equation we find immediately that if a solution exists, then 
\begin{equation}
  \label{eq:power-law-X}
  X=X_{0}=\frac{2-\gamma }{4-3\gamma }=const.
\end{equation}
Expressing $\varepsilon$ in terms of $p$ from (\ref{eq:gamma}) and substituting
(\ref{eq:new-lagrangian}) with $X$ given by (\ref{eq:power-law-X}) into the
master Eq. (\ref{eq:master}), we get a simple equation for $f$, which is  
solved by 
\begin{equation}
  \label{eq:power-law-f}
  f\left( \varphi \right) =\frac{4}{9}\frac{\left( 4-3\gamma \right)}{\gamma
    ^{2}}\frac{1}{\left(\varphi -\varphi_*\right) ^{2}}.
\end{equation}
Therefore, a model with Lagrangian (\ref{eq:new-lagrangian}), with
$f\left(\varphi\right)$ given by Eq. (\ref{eq:power-law-f}), has an
attractor solution which describes power law expansion,
\begin{equation}
  \label{eq:power-law-a}
  a\left(t\right) \varpropto t^{\frac{2}{3\gamma}}.
\end{equation}
If $0<\gamma <2/3$ then this solution describes the usual power law inflation.
If one takes a negative value of $\gamma$ then one gets pole-like
super-inflation in \textit{Einstein frame}. However, this pole-like inflation
has a ``graceful exit problem'' which is very similar to the one of the pre-big
bang scenario  \cite{PBB}. We were unable to find a simple solution to this 
problem and we doubt that such a solution can be meaningfully discussed within 
the effective field Lagrangian formalism considered here. Therefore we
think that pole-like inflation does not help toward bringing a solution of
the main cosmological problems. In distinction from pole-like inflation, in the
model of power law inflation the graceful exit problem can be easily solved
if in some range of $\varphi$ the function $f$ is modified in an obvious way.

A natural generalization of the  Lagrangian (\ref{eq:new-lagrangian}),
\begin{equation}
  \label{eq:general-lagrangian}
  p=f\left( \varphi \right) g\left(X\right),
\end{equation}
where $g$ is a rather arbitrary function of $X$, opens the possibility to
realize power law inflation in a wide class of theories. Actually, taking
the function $f$ to be
\begin{equation}
  \label{eq:generalized-f}
  f\left( \varphi\right) =-\frac{4}{9}\frac{\left(1-\gamma\right)}{\gamma
    ^{2}}\frac{2X_{0}}{g\left( X_{0}\right)}\frac{1}{\left(\varphi-\varphi
      _*\right)^2},
\end{equation}
where $X_{0}$ is a solution of the equation 
\begin{equation}
  \label{eq:solX}
  2X\frac{\partial\ln g}{\partial X}=\frac{\gamma}{\gamma-1},
\end{equation}
one can easily verify that power law inflation (\ref{eq:power-law-a}) is a
solution of the corresponding theory.


\section{Stability}
\setcounter{equation}{0}

A detailed study of the spectrum of quantum perturbations in our slow-roll 
\textit{k}-inflation scenario will be done in a forthcoming publication. We
shall only discuss here a necessary condition for the \textit{stability} of our
models with respect to arbitrary, high-frequency scalar perturbations. The
equation for the canonical ``quantization variable'' $v$  describing the  
collective 
metric and scalar field perturbations in the case of the action
(\ref{eq:einstein-action}) can be written down in the  standard way \cite{MFB}
and takes the form \cite{GM99}
\begin{equation}
  \label{eq:perturbations}
  v''-c_s^2\nabla ^2 v-\frac{z''}{z}v=0 ,
\end{equation}
where the only relevant piece of information for our stability analysis is 
the appearance of the ``speed of sound'' 
\begin{equation}
  \label{eq:sound-velocities}
  c_{s}^{2}=\frac{p_{,X}}{\varepsilon_{,X}}=\frac{p_{,X}}{p_{,X}+2Xp_{,XX}}
\end{equation}
in front of the Laplacian. Here a comma denotes a partial derivative with
respect to $X.$ It is clear that if $c_{s}^{2}$ is negative then the model
is absolutely unstable. The increment of instability is inversely
proportional to the wavelength of the perturbations, and therefore the
background models for which $c_{s}^{2}<0$ are violently unstable and do not have 
any physical significance.

This stability requirement ($c_{s}^{2}=p_{,X}/\varepsilon _{,X}>0$) is non
trivial within our scenario, because, for instance, in slow-roll 
\textit{k}-inflation in the zeroth-order slow-roll approximation the
inflationary attractors are defined by $p_{,X}=0$, and therefore
$c_{s}^{2}=0$. However, as discussed in Section \ref{sec:slow-roll-inflation},
in the post-slow-roll approximation, with $X=X_{\rm sr}+\delta X$, $p_{,X}$
does not vanish. To first order in $\delta X$ we can write
$p_{,X}\simeq p_{,XX}^{\rm sr}\delta X$. Using the second equation
(\ref{eq:sound-velocities}) we get 
\begin{equation}
  c_{s}^{2}\simeq \frac{\delta X}{2X_{\rm sr}}.
\end{equation}
Therefore stability requires that $\delta X>0$, i.e. that on the equation of
state graphs of Figs. \ref{fig:state} and \ref{fig:model-state}, the $%
\varphi $-gradients of $p(\varphi,X)$ be such that they displace the real,
non-adiabatic, slow-roll attractor \textit{beyond} the $p=-\varepsilon $
line (``beyond'' meaning here ``further away'' as one runs along the $%
p=f(\varepsilon ,\varphi )$ graph following the natural $X$ parametrization). 
These stable stretches of the $%
(\varepsilon ,p)$ graphs are labelled $s$ in Fig. \ref{fig:state}. They are
also the stretches where the slope $(dp/d\varepsilon)$ is positive (as is
clear from Eq. (\ref{eq:sound-velocities}) which says that the velocity of
sound is given by the usual formula $c_{s}^{2}=dp/d\varepsilon$, when
keeping $\varphi$ fixed).

Let us now discuss the simple model (\ref{eq:simplest-p-expansion}). For
this model we have computed $\delta X/X_{\rm sr}$, and we can therefore assess
under what conditions \textit{k}-inflation will be stable under scalar
perturbations. We see from Eq. (\ref{eq:slow-roll-parameter}) that the
conditions can be expressed in two (equivalent) forms: (i) the energy
density $\varepsilon=\sqrt{\overline{K}}$ must \textit{decrease} during the
slow-roll of $\varphi$, or (ii) $-\sigma\overline{K}_{,\varphi}=\sigma
K_{,\varphi}$ must be positive. The satisfaction of this condition is very
natural within the intuitive picture we have in mind: namely, starting at
some high ($\sim$ Planckian) energy density, i.e. a large \textit{negative}
value of $K(\varphi)=-\varepsilon^2$, and then letting $\varphi$ evolve
toward the weak-field coupling domain where $K(\varphi)$ vanishes before
becoming positive. During the slow-roll phase (with $K(\varphi)<0$), it is
natural (and even necessary if $K(\varphi)$ is monotonic) to have a
decreasing $\overline{K}=-K$.

One consequence applicable to a general model $p(\varphi,X)$ is that
slow-roll implies a small value $|c_{s}^{2}|\ll 1$. It is interesting to ask
for which (non slow-roll) models one can have both a continued
\textit{k}-inflation and a constant speed of sound of order unity. Let us
consider for that purpose power law inflation. In the case where the
Lagrangian takes the form (\ref{eq:new-lagrangian}) the speed of sound during
the inflationary stage is 
\begin{equation}
  \label{eq:power-law-cssq}
  c_{s}^{2}=\frac{\gamma}{8-3\gamma}. 
\end{equation}
If we restrict ourselves to the inflationary range $0<\gamma<2/3$ this speed can 
not exceed $c_{s}^{2}=1/9.$ The smaller
values of $\gamma $ correspond to very fast (nearly exponential) expansion
and small speed of sound in complete agreement with our analysis of slow-roll 
\textit{k}-inflation. Pole-like inflation is violently unstable in this
model.

However, if we consider more general Lagrangians (\ref{eq:general-lagrangian})
we can avoid these restrictions. Actually in this case the speed of sound
during the inflationary stage is given by the expression 
\begin{equation}
  c_{s}^{2}=\frac{g_{,X}\left( X_{0}\right)}{g_{,X}\left(X_{0}\right)
    +2X_{0}\,g_{,XX}\left(X_{0}\right)},
\end{equation}
where $X_{0}$ is the solution of equation (\ref{eq:solX}). The necessary
conditions for power law inflation given in Section
\ref{sec:power-law-kinflation} imply that $g_{,X}\left( X_{0}\right) \not= 0$ 
and do not involve any restrictions on the second
derivative $g_{,XX}\left( X_{0}\right)$. Therefore, for a power law
inflationary stage with any a priori given value of the parameter $\gamma$, one 
can always find a 
corresponding function $g\left(X\right)$ to arrange any required speed of
sound. Note that it follows from here that one can also easily build a theory 
with pole-like inflation which is stable with respect to scalar perturbations.


\section{Exit Mechanisms}
\label{sec:exit-mechanisms}
\setcounter{equation}{0}

In the simple model of Section \ref{sec:slow-roll-inflation} (and in the  
normalization $L(\varphi)=1$) the total number of
inflationary $e$-folds is (considering for definiteness that $\varphi$
decreases during slow-roll) 
\begin{equation}
  N_{\rm inf}=\frac{1}{2}\int\limits_{\varphi_{\rm end}}^{\varphi_{\rm in}}
  \sqrt{\overline{K}(\varphi)}\,d\varphi.
\end{equation}
Here $\varphi_{\rm in}$ is the initial value of $\varphi$, and
$\varphi_{\rm end}$ the end of slow-roll, i.e. the value of $\varphi$ where
$\delta X/X$ becomes of order unity, i.e. (from Eq.
(\ref{eq:slow-roll-parameter})), such that 
\begin{equation}
  \label{eq:slow-roll-end}
  \overline{K}^{-3/2}(\varphi_{\rm end})\overline{K}^{\prime}(\varphi_{\rm 
end})\sim 1.
\end{equation}

We shall not investigate here the problem of the choice of initial
conditions within our model. We shall assume that some large parameter is
present (or at least possible) in the problem and allows $N_{\rm inf}$ to be
larger that 60 or so. [One simple possibility would be a function $K(\varphi
)$ of the type of Eq. (\ref{eq:FMS-K}) which levels off to a negative
constant when $\varphi $ increases. All the couples $(\varphi,X)$ where
$X\sim 1$ (in Planck units) and $\varphi $ is arbitrary large lead to an
energy density of order $1$. A random initial condition with $\varepsilon
(\varphi_{\rm in},X_{\rm in})\sim 1$ could have an arbitrary large $\varphi$.]
Assuming this we note not only that our mechanism contains a natural exit
from inflation (because of the evolution of $K$ and its final change of
sign), but that this exit is generically expected to take place within a small
number of $e$-folds. Indeed, condition (\ref{eq:slow-roll-end}) signalling the
end of slow-roll \textit{k}-inflation can be rewritten (using Eqs. (\ref
{eq:X-slow-roll})-(\ref{eq:H-slow-roll})) as $[\dot{K}/(KH)]_{\rm end}\sim 1$,
which means that $K$ changes by 100\% in a Hubble time, around the time of
exit of \textit{k}-inflation. One therefore expects that one can approximately
match slow-roll \textit{k}-inflation with a post-inflationary phase where
$K(\varphi )$ has become positive.

We think that, in most cases, this way of exiting \textit{k}-inflation
provides a naturally graceful exit. Indeed, it is clear from Fig.
\ref{fig:model-state} that, after the transition to the $K>0$
(``weak-coupling'') branch, the cosmological evolution will quickly be
attracted toward an approximate $p\approx\varepsilon$ equation of state. The
corresponding expansion was discussed in Eq. (\ref{eq:vacuum-attractor}) and
corresponds to a very fast decrease of the energy density:
$\varepsilon_{\varphi}\propto a^{-6}$. As this decrease is much faster than
the decay of the energy density in radiation ($\varepsilon_{\rm rad}\propto
a^{-4}$) [and in non relativistic matter ($\varepsilon_{\rm matter}\propto
a^{-3}$) if any is present], even small traces of the latter forms of energy
present at the end of \textit{k}-inflation will ultimately dominate the
expansion. The situation is very similar to what has been recently discussed
in Ref. \cite{PV98}. As we have in mind that, in our model, the scalar
$\varphi$ could be the dilaton (or a moduli), i.e. a field which modifies the
coupling constants of all the other matter fields, we expect that the nearly
uniform time variation of $\varphi$ during \textit{k}-inflation will generate
quantum particles at a uniform spacetime rate. During \textit{k}-inflation the
produced particles are constantly diluted by the fast expansion and are not
expected to cause a strong back reaction, but the particles produced in the
last $e$-fold of \textit{k}-inflation should be sufficiently
numerous to dominate soon the expansion. Even without this assumption (that
the couplings of $\varphi$ are efficient in producing particles), the mere
effect of the variable gravitational coupling (at the end of inflation) is
sufficient to create any scalar particle with energy density (at birth)
$\sim 10^{-2}H_{\rm end}^4$ \cite{Ford} \cite{DV96}.

To discuss more precisely the evolution of the inflaton after it exits from
\textit{k}-inflation, i.e. when $K(\varphi)$ has become positive, and the
higher-derivative term $L\,X^2$ has become negligible, it is convenient to
introduce the canonical scalar field $\phi=\int d\varphi \sqrt{K(\varphi)}$.
In terms of $\phi$ the equation of motion reads $\ddot{\phi}+3H\dot{\phi}%
=a^{-3}d(a^3\dot{\phi})/dt\approx 0$, with $H=\sqrt{\varepsilon_\phi+%
\varepsilon_{\rm rad}}$. Hence $a^3\dot{\phi}\simeq c=\mathrm{const}$ and
$\phi$ evolves according to $\phi(t)\approx c\int dt\,a^{-3}$. In a first phase
after \textit{k}-inflation $\varepsilon_\phi$ probably dominates over
$\varepsilon_{\rm rad}$ and the evolution follows the $p=\varepsilon$ attractor
solution Eq. (\ref{eq:vacuum-attractor}). During this initial phase
$a(t)\propto t^{1/3}$ so that $\phi(t)$ drifts logarithmically:
$\phi(t)\propto \int_{t_{\rm end}}^t dt/t=\ln(t/t_{\rm end})$. Later,
$\varepsilon_{\rm rad}\propto a^{-4}$ will take over $\varepsilon_\phi\propto
a^{-6}$ and the evolution will become radiation dominated. In this second
phase, $a(t)\propto t^{1/2}$ and $\phi(t)\simeq c\int^t dt a^{-3}\propto
\int^t dt\,t^{-3/2}$ stops drifting logarithmically to converge toward some
final value $\phi_f$: $\phi(t)\approx \phi_f-c^{\prime}t^{-1/2}$.

Note that, in this generic exit mechanism, the final value $\varphi_f$
of the original field (corresponding to the final value $\phi_f$ of the
canonical field) is arbitrary. Therefore, if $\varphi$ is the dilaton or a
moduli our \textit{k}-inflationary mechanism does not, by itself, provide a
mechanism for fixing $\varphi$ to a particular value. To do that one must
appeal either to the presence, at low-energies, of a $\varphi$-dependent
potential energy term $V(\varphi)$ (which may have been negligible at high
energies), or to a non-trivial structure of couplings to matter \cite{DP94}.
We wish, however, to point out that some variants of our general model
can also provide another way of fixing the end location of $\varphi$ very near
a particular value. Indeed, if the kinetic function $K(\varphi)$ (of the
original $\varphi$ variable) happens to have a pole singularity $%
K(\varphi)\propto (\varphi-\varphi_{p})^{-\alpha}$ with $\alpha\geq 2$ in
the positive-$K$ domain the corresponding canonical field $\phi=\int d\varphi 
\sqrt{K(\varphi)}$ diverges when $\varphi\to\varphi_{p}$. Therefore, if
this pole singularity is in the way of the evolution of $\varphi$ after the
exit from \textit{k}-inflation (e.g., if $\varphi_{p}<\varphi_{\rm end}
<\varphi_{\rm in}$ in the case where $\dot{\varphi}<0$), the typically large
logarithmic drift of the canonical field $\phi$ during the first phase after
exit, $\Delta \phi\simeq c\int_{t_{\rm end}}^{t_{\rm rad}}dt\,a^{-3}\propto
\ln(t_{\rm rad}/t_{\rm end})$ (where $t_{\rm rad}$ denotes the beginning of
radiation domination), will mean that the original field $\varphi$ will end up 
very near $\varphi_p$.

Evidently, this mechanism assumes that, except for $K(\varphi)$, all the
functions describing the coupling of $\varphi$ to matter (like the one
giving the $\varphi$-dependence of the gauge couplings) are regular (or, at
least, less singular) at $\varphi=\varphi_p$. Then, in the notation of
Ref. \cite{DP94}, the observable coupling strength of $\varphi$ to matter
$\alpha_A(\varphi)=d\ln m_A(\phi)/d\phi=(K(\varphi))^{-1/2}d\ln
m_A(\varphi)/d\varphi$ is driven near zero by our mechanism.


\section{Conclusions}
\setcounter{equation}{0}

We have pointed out that a general class of higher-order scalar kinetic
terms $p(\varphi,\nabla \varphi)=\frac{1}{2}K(\varphi)(\nabla\varphi
)^{2}+\frac{1}{4}L(\varphi)(\nabla\varphi)^{4}+\cdots$ can drive an
inflationary evolution starting from rather arbitrary initial conditions.
Under not very restrictive conditions on the $\varphi $-dependence of the
kinetic term $p(\varphi,\nabla\varphi)$ the early cosmological evolution
will be attracted toward a slow-roll kinetically driven inflationary stage.
In a large class of models this slow-roll behaviour has the following
attractive features: (i) it drives the evolution from an initial
high-curvature phase down to lower curvatures while dilating space in a
quasi-exponential or power law manner, (ii) it is stable under
(high-frequency) scalar perturbations, and (iii) it contains a natural exit
mechanism because of the $\varphi$-dependence of the kinetic terms.

We have briefly discussed the exit of kinetically driven inflation and found
that it seems to be naturally ``graceful'' in lending itself to a smooth
transition toward a stage dominated by the radiation produced (either
through the $\varphi$-dependence, or through purely gravitational effects)
at the end of slow-roll. We have also pointed out that the presence of
pole-like singularities in the $\varphi$-dependence of the kinetic terms can
have some useful consequences: (i) a $\varphi$-dependence of the form $\sim
(\varphi-\varphi_*)^{-2}\tilde{p}(\nabla\varphi)$ in the initial field domain
can ensure a nearly constant speed of sound of order unity, while (ii) a
pole-like singularity in the coefficient $(\nabla\varphi)^2$ in the final
field domain can help to fix the value of $\varphi$ to a specific value.

We leave to future work the investigation of general important issues: the
choice of (a measure on the) initial conditions, and the computation of the
perturbation spectra generated by this new type of inflation. We are aware
that the compatibility of the latter issues with observations will probably 
necessitate the presence of some small (or large) dimensionless parameters in
$p(\varphi,\nabla\varphi)$.

In this work, we have used the structure of the effective action in string
theory as a partial motivation for considering higher-order kinetic terms
$p(\varphi,\nabla\varphi)$. Having found that such terms can generically
drive an inflationary behaviour, we hope that our mechanism might be useful
in suggesting new ways in which the dilaton and moduli fields of string
theory might be compatible with inflation.


\section*{Acknowledgements}
The work of V.M. and C.A.P. was supported in part by the
\textit{Sonderforschungsbereich 375-95 der Deutschen Forschungsgemeinschaft}. 
T.D.'s work was partly supported by the NASA grant NAS8-39225 to Gravity Probe 
B. C.A.P. thanks the \textit{Institut des Hautes \'Etudes Scientifiques} for its
kind hospitality. 



\end{document}